\documentclass[12pt,a4paper,dvips]{article}
\usepackage{a4p}
\usepackage{cite,mcite}
\usepackage{graphicx}
\usepackage{physics }
\usepackage{l3_title,ifthen}
\usepackage{amsmath,rotating,color,amssymb}

\journalname{Eur. Phys. Jour. C}

\date{November 12, 2003}
\preprint{2003-072}
\newlength{\capindent}
\newlength{\capwidth}
\newlength{\figwidth}
\newcommand{\icaption}[2][!*!,!]{\hspace*{\capindent}%
  \begin{minipage}{\capwidth}
    \ifthenelse{\equal{#1}{!*!,!}}%
      {\caption{#2}}%
      {\caption[#1]{#2}}
  \end{minipage}}
\newcommand{\pho}{\phantom{0}}

\newcommand{\ww}{\Wp\Wm}
\newcommand{\lnqq}{\ensuremath{\ell\nu\qqbar}}
\newcommand{\enqq}{\ensuremath{\text{e}\nu\text{q}\bar{\text{q}}}}
\newcommand{\mnqq}{\ensuremath{\mu\nu\text{q}\bar{\text{q}}}}
\newcommand{\tnqq}{\ensuremath{\tau\nu\text{q}\bar{\text{q}}}}
\newcommand{\qqqq}{\ensuremath{\text{q}\bar{\text{q}}\text{q}\bar{\text{q}}}}

\newcommand{\fm}{\ensuremath{f_{-}}}
\newcommand{\fp}{\ensuremath{f_{+}}}
\newcommand{\fpm}{\ensuremath{f_{\pm}}}
\newcommand{\fo}{\ensuremath{f_{0}}}

\newcommand{\tlep}{{\ensuremath{\theta^*_\ell}}}
\newcommand{\thad}{{\ensuremath{\theta^*_{\text{q}}}}}

\newcommand{\costlep}{\ensuremath{\cos\tlep}}
\newcommand{\costhad}{\ensuremath{|\cos\thad|}}

\newcommand{\thetawm}{\ensuremath{\Theta_\Wm}}
\newcommand{\cosT}{\ensuremath{\cos\thetawm}}
\newcommand{\pmone}{\ensuremath{\pm 1}}
\newcommand{\dphi}{\ensuremath{\left|\Delta\phi\right|}}

\begin{document}
\begin{titlepage}
\title{Study of Spin and Decay-Plane Correlations \\of W Bosons in the e$^+$e$^-\rightarrow$ W$^+$W$^-$ Process at LEP}
\author{The L3 Collaboration}
%
%
\begin{abstract}
Data collected at LEP at centre-of-mass energies $\rts{} = 189-209
\GeV{}$ are used to study correlations of the spin of W bosons using
$\epem\ra\ww\ra\lnqq{}$ events. Spin correlations are favoured by
data, and found to agree with the Standard Model predictions.  In
addition, correlations between the W-boson decay planes are studied in
$\epem{}\ra{}\ww{}\ra{}\lnqq{}$ and $\epem{}\ra{}\ww{}\ra{}\qqqq{}$
events.  Decay-plane correlations are measured to be consistent with
the Standard Model predictions.
\end{abstract}

\submitted

\end{titlepage}
%
%

\section*{Introduction}

The study of the properties of the W boson constitutes one of the main
physics goals of the LEP experiments.  We have previously reported on
the measurement of the helicity fractions of W bosons in
$\epem\ra\ww\ra\lnqq{}$ events, with $\ell$ denoting either an
electron or a muon, using data collected at LEP at centre-of-mass
energies, \rts{}, up to 209 \GeV{}~\cite{wpol}. Measurements of
W-boson polarisation were also reported in the framework of a
spin-density matrix analysis~\cite{opal}. The same data sample
is used here to study correlations of the spin of W bosons as well as
correlations between the decay planes of the two W bosons.  This last
study also includes $\epem\ra\ww\ra\qqqq{}$ events.  Some correlations
of the spin of W bosons in the $\epem{}\ra{}\ww{}$ process are
expected in the Standard Model and are accessible with the size of the
data sample collected at LEP. Weak decay plane correlations are also
expected.

Studies of triple-gauge-couplings of the W boson~\cite{tgc}, which
rely on the analysis of the full differential cross section of the
$\epem{}\ra{}\ww{}$ process, have an implicit sensitivity to these
correlations. Their results are presented in terms of couplings
describing the Lorentz-invariant $\gamma$WW and ZWW vertices. The aim
of the analysis presented in this Article, as a natural extension of
the work of Reference~\citen{wpol}, is a direct and model independent
measurement of the spin and decay-plane correlations. This enables a
specific test of the corresponding Standard Model predictions, as
suggested in References~\citen{bilenky} and~\citen{kane1} for the spin
and decay-plane correlations, respectively.  Large deviations from the
Standard Model predictions, possibly outside the reach of LEP, would
suggest anomalous mechanisms for W-boson pair-production in $\epem$
collisions such as one-loop contributions from new particles or in
general effects from any model with a symmetry breaking mechanism
which affects interactions of longitudinally-polarised W bosons.  To
date, no such direct measurement was reported and this study
introduces a new technique which could be exploited at future
high-luminosity lepton colliders.

W-boson spin correlations are measured in
$\epem{}\ra{}\ww{}\ra{}\lnqq{}$ events by tagging the helicity of the
W boson which decays into hadrons and measuring the helicity of the W
boson which decays into leptons. We consider two subsamples of W-boson
pairs: the first is enriched in events where \Wm{} bosons decaying
into hadrons have a helicity $\lambda_{\Wm{}} = \pmone{}$ and the
second is depleted of these events
\footnote{The charge conjugate state \Wp{} is also included throughout
this Article. CP conservation is assumed, as verified in W-boson
polarisation studies~\cite{wpol,opal}.}.  A difference in the helicity
compositions of the W bosons decaying into leptons for the two
subsamples would indicate the presence of W-boson spin correlations. 

The helicity fractions of the \Wm{} boson decaying into leptons are
obtained in a model independent approach from the shape of the
distribution of the polar decay angle, \tlep{}, between the charged
lepton and the \Wm{}-boson flight direction in the \Wm{}-boson rest
frame.  The differential distribution of leptonic \Wm{}-boson decays
is:
\begin{equation}
\frac{1}{N}\frac{\mathrm{d}N}{\mathrm{d}\cos\tlep{}} = \fm{}
\frac{3}{8}~(1+\cos\tlep{})^{2} + \fp{}
\frac{3}{8}~(1-\cos\tlep{})^{2} + \fo{} \frac{3}{4} \sin^{2} \tlep{},
\end{equation}
where \fm{}, \fp{} and \fo{} represent the fractions of \Wm{} bosons
in the $-1$, $+1$ and 0 helicity states, respectively.  Assuming CP
invariance, these equal the fractions of the corresponding helicity
states $+1$, $-1$ and 0 of the \Wp{} boson.  For hadronic W-boson
decays, considering only the absolute value of the cosine of the polar
decay angle, \costhad{}, the differential distribution is:
\begin{equation}
\frac{1}{N}\frac{\mathrm{d}N}{\mathrm{d}\costhad} = \fpm{} \frac{3}{4}
(1+\costhad{}^{2}) + \fo{} \frac{3}{2} (1-\costhad{}^2),
\end{equation}
with \fpm{}=\fp{}+\fm{}. It is worthwhile to remark that the only
hypothesis in the derivation of Equations~1 and~2 is the decay of a
unit-spin W boson into fermions.

The values of  \fm{}, \fp{} and \fo{} are a function of $\sqrt{s}$ and
the \Wm{}-boson production angle with respect to the incoming e$^-$,
\cosT{}. The numerical values corresponding to the data sample
investigated in this Article are therefore dependent on the integrated
luminosity collected at each centre-of-mass energy at which the LEP
collider was operated.

\begin{figure}[htbp]
\begin{center}
    \includegraphics[height=17.5cm]{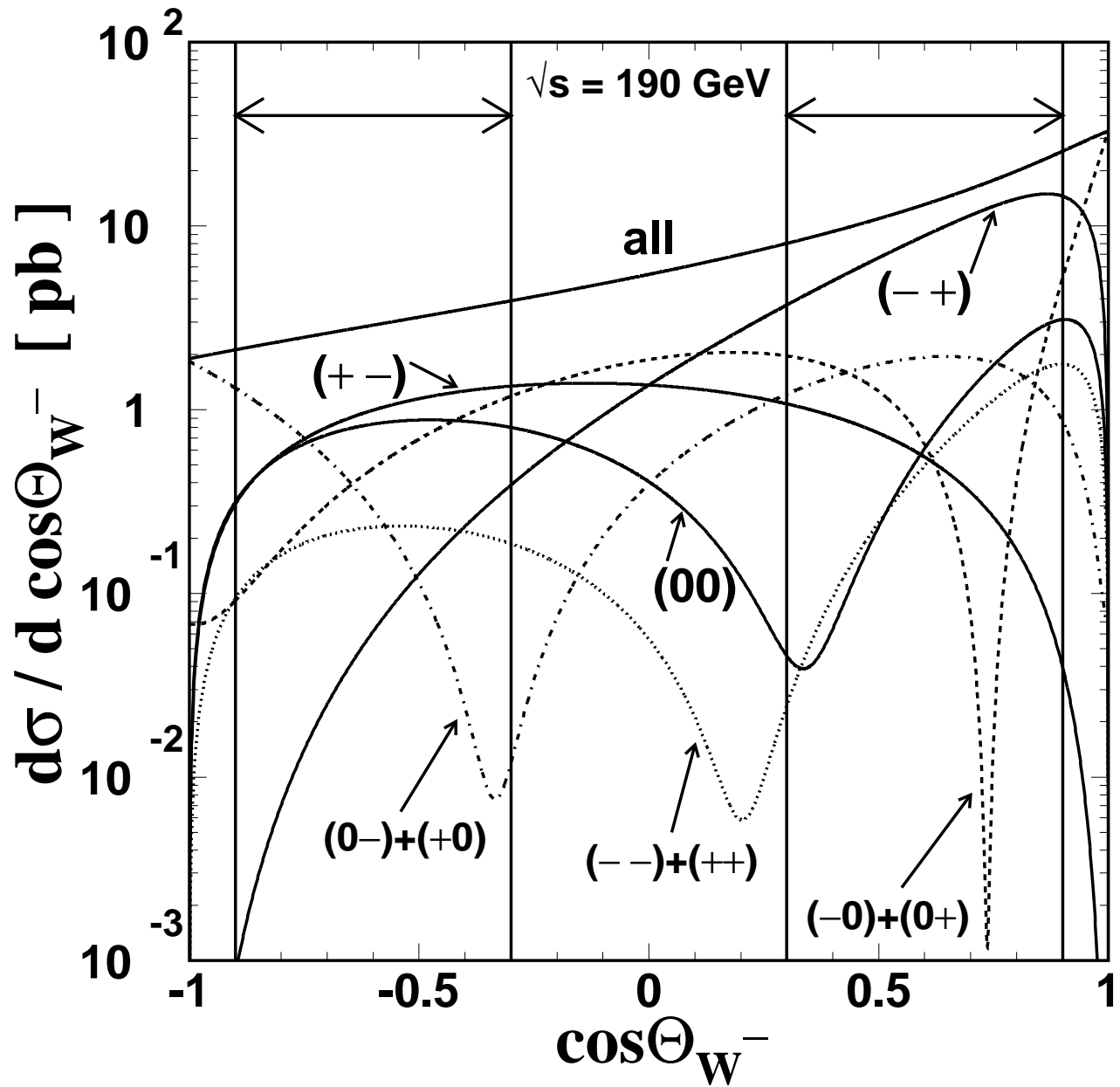}
\end{center}
\caption{Differential cross section for pair production of polarised W bosons 
at \rts{} = 190 \gev{}  averaged over initial electron polarisations. 
The \Wm{} and \Wp{} helicities in the \epem{} centre-of-mass frame are given in parentheses. 
The intervals used for the analysis of W-boson spin correlations are indicated by the arrows.
\label{fig:hagiwara}}
\end{figure}

The fractions of helicity combinations
$(\lambda_{\Wm{}},\lambda_{\Wp{}})$ depend strongly on \cosT{}. As an
example, Figure~\ref{fig:hagiwara} shows the results of a
leading-order analytical calculation~\cite{hagiwara}. This property is
used to select intervals in which a particular helicity combination is
enriched and thus spin correlations are more significant.  Two
intervals are considered: the forward bin, $0.3 < \cosT{} < 0.9$,
where the fraction of the helicity combination
$(\lambda_{\Wm{}},\lambda_{\Wp{}}) = (-1,+1)$ is increased to about
68\% of all W-boson pairs, compared to an average value of 49\% over
the whole \cosT{} range; and the backward bin, $-0.9 < \cosT{} <
-0.3$, where the fraction of the helicity combination
$(\lambda_{\Wm{}},\lambda_{\Wp{}}) = (0,0)$ is increased to about
26\%, compared to an average value of 8\%.

\begin{figure}[htbp]
\begin{center}
    \includegraphics[height=14cm]{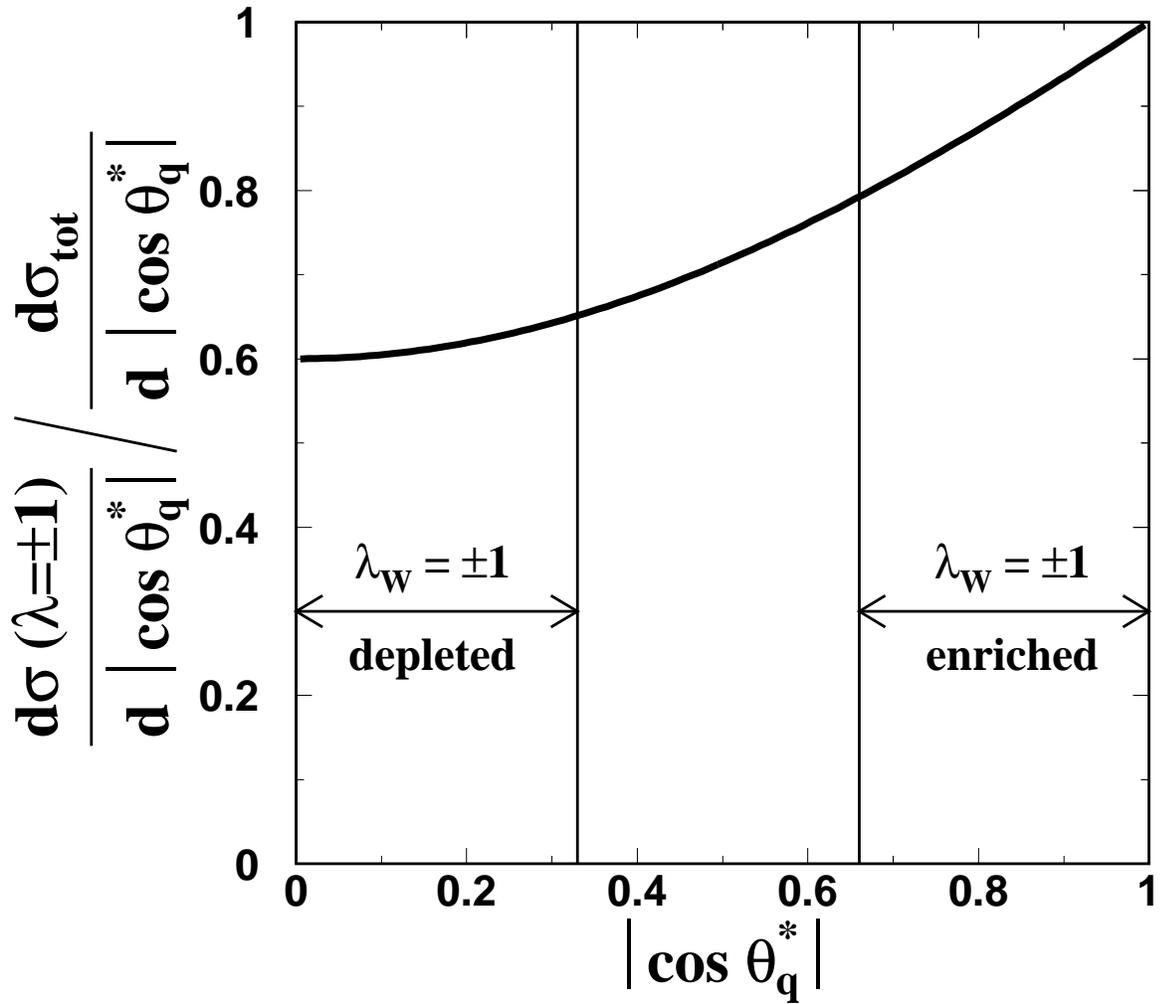}
\end{center}
\caption{Standard Model predictions for the relative contribution of
  the helicity states \pmone{} to the $\epem\ra\ww\ra\lnqq{}$
  differential cross section 
as a function of \costhad{}. The intervals used for the analysis of W-boson spin correlations are indicated.
\label{fig:funhad}}
\end{figure}

To tag the helicity of the W boson which decays into hadrons, cuts on
$\costhad{}$ are used.  According to Equation 2, for small values of
\costhad{}, the sample is depleted of helicity \pmone{} states, while
for large values of \costhad{} the sample is enriched in helicity
\pmone{} states, as shown in Figure~\ref{fig:funhad}.  The interval
$\costhad{} < 0.33$ is chosen for the $\lambda_{\Wm{}} = \pmone{}$
depleted sample and the interval $\costhad > 0.66$ for the
$\lambda_{\Wm{}} = \pmone{}$ enriched sample. The helicity fractions
of the \Wm{} bosons decaying into leptons are obtained from a fit of
Equation 1 to the distribution of \costlep{}~\cite{thesis}.

Decay-plane correlations are studied in both
$\epem{}\ra{}\ww{}\ra{}\lnqq{}$ and $\epem\ra\ww\ra\qqqq{}$ events
using the absolute value of the angle, \dphi{}, between the planes
defined by the decay products of the two W bosons in the rest frame of
the W-boson pair.  The strength of the correlation is measured by the
parameter $D$ of the differential distribution~\cite{kane}:
\begin{equation}
\frac{1}{N}\frac{\mathrm{d}N}{\mathrm{d}\dphi{}} = 1 + D \cos 2\dphi.
\end{equation}
The experimental distribution of \dphi{} is fitted to obtain the
parameter $D$~\cite{thesis}.  The expected value of $D$ in the
Standard Model, for the \rts{} range under investigation, is estimated
with a leading-order analytical calculation~\cite{kane} to vary from 0.021 at
$\sqrt{s}=189\gev$ to 0.015 at $\sqrt{s}=209\gev$, with a
luminosity-weighted average of 0.018.

For both analyses, prior to the fits, the distributions of the
measured angles are corrected by means of an efficiency correction
function which accounts for selection efficiencies, migration effects
and the presence of initial state radiation. This correction function
is obtained from large Monte Carlo samples as the ratio of the angular
distributions for simulated and generated events.

\section*{Data and Monte Carlo Samples}
The analysis is based on $629.3\,\pb{}$ of data collected with the L3
detector~\cite{l3detector} at $\rts{} =  189-209 \gev{}$, as detailed
in Table~\ref{tab:table1}, corresponding to a luminosity-weighted
average $\sqrt{s}$ value of $197.9\gev{}$.

Signal Monte Carlo events are generated using the
KORALW~\cite{KORALW1,*KORALW2} program for the
$\epem{}\ra{}\Wp{}\Wm{}\ra{}\enqq{}, \mnqq{}$ and $\qqqq{}$ processes.
The Standard Model predictions for the W-boson spin correlations and
the decay-plane correlations are obtained from the distributions
generated at different values of \rts{}, combined according to the
collected luminosity.  Background processes are generated using KORALW
for all other final states of W-boson pair-production,
KK2f~\cite{KK2F} for the $\epem\ra{}\qqbar(\gamma)$ process, and
PYTHIA~\cite{PYTHIA} for the $\epem\ra{}\text{ZZ}$ process.  For
studies of systematic effects, signal events are also generated using
the EEWW~\cite{EEWW}, YFSWW \cite{YFSWW3} and
EXCALIBUR~\cite{excalibur} programs.

To test the consistency of the predictions for W-boson spin and
decay-plane correlations, large samples of signal events are generated
using the EEWW and YFSWW Monte Carlo programs.  These differ from
KORALW in the level and implementation of $\mathcal{O}(\alpha)$
corrections.  The predictions given by the three programs for the
strength of the correlations are in agreement with each other, within
their own statistical uncertainties.

The predicted Standard Model value of the parameter $D$, which describes the
decay-plane correlations is $D = 0.010 \pm 0.002$, as obtained with the
KORALW Monte Carlo. The uncertainty is statistical.
This differs from the value $D = 0.018 $ obtained by an analytical
calculation~\cite{kane} due to the inclusion of radiative effects in
the Monte Carlo.

The L3 detector response is simulated with the GEANT~\cite{geant} and
GHEISHA~\cite{gheisha} packages.  Detector inefficiencies, as
monitored during the data taking periods, are included.

After detector simulation, the average resolution on $\cosT{}$ is found to be
0.06 while the resolution on $\cos\theta^*$ is found to be 0.10 for
electrons,  0.11 for muons and 0.14 for hadrons. The resolution on $\dphi{}$ is
$7.5^\circ$ for semi-leptonic events and $10.0^\circ$ for hadronic events.

\section*{Event Selection}
The selection of $\epem\ra\ww\ra\lnqq{}$ events is the same as for the
study of the W-boson polarisation~\cite{wpol}, however, the
low-statistics data sample collected at $\sqrt{s}=183\GeV$ is not
considered here.  The numbers of selected events are listed in
Table~\ref{tab:table1} for different values of \rts{}.  In total, 1861
events are selected with an average efficiency of 67.6\% and an
average purity of 96.6\%, which are only slightly dependent on $\sqrt{s}$.
The contamination from the $\epem{}\ra{}\Wp{}\Wm{}\ra{}\tnqq{}$ and
$\epem\ra{}\qqbar(\gamma)$ processes is 2.3\% and 1.1\%, respectively.

\begin{figure}[htbp]
  \begin{center}
    \includegraphics[width=8cm]{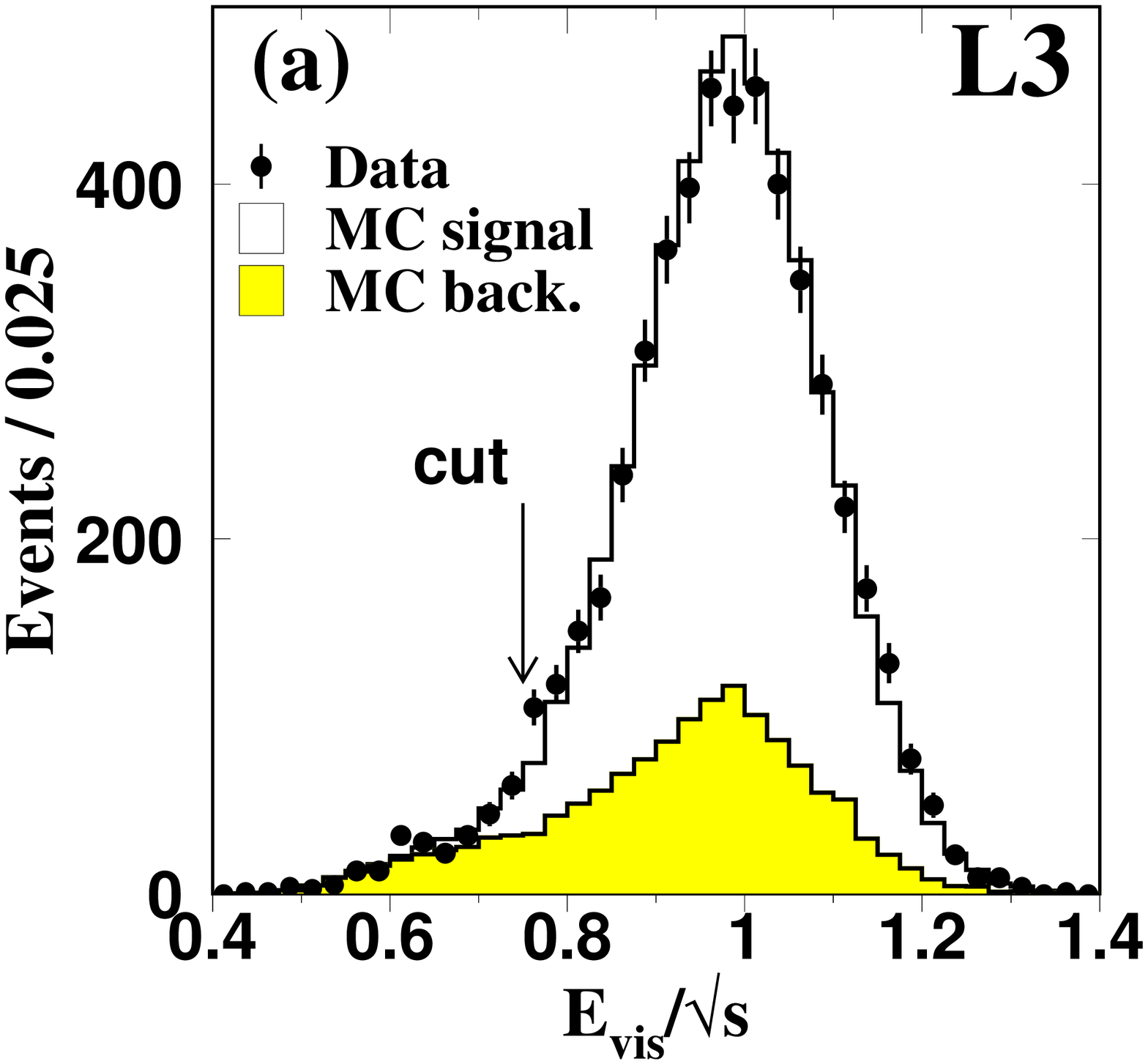}
    \includegraphics[width=8cm]{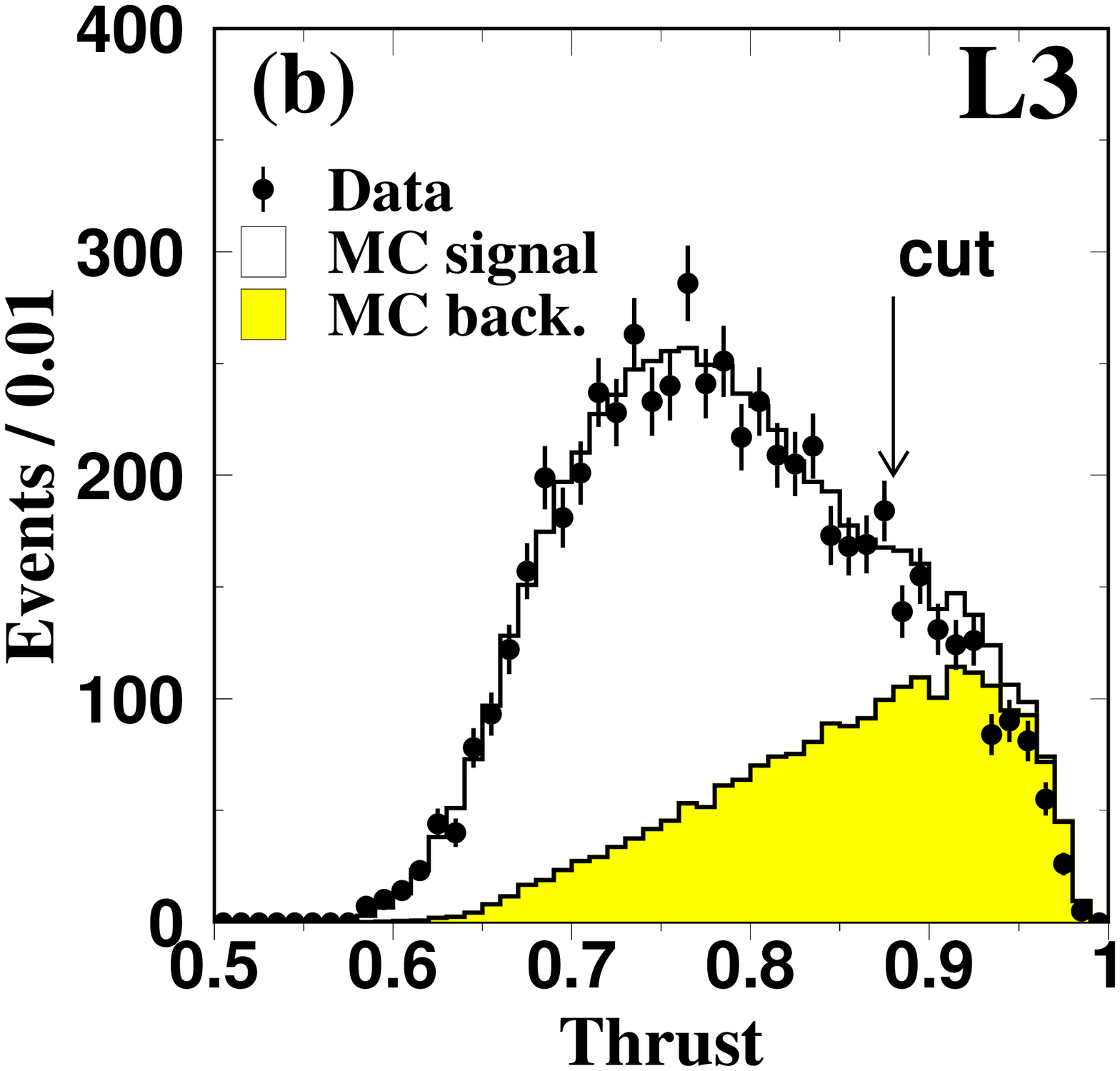}\\
    \includegraphics[width=8cm]{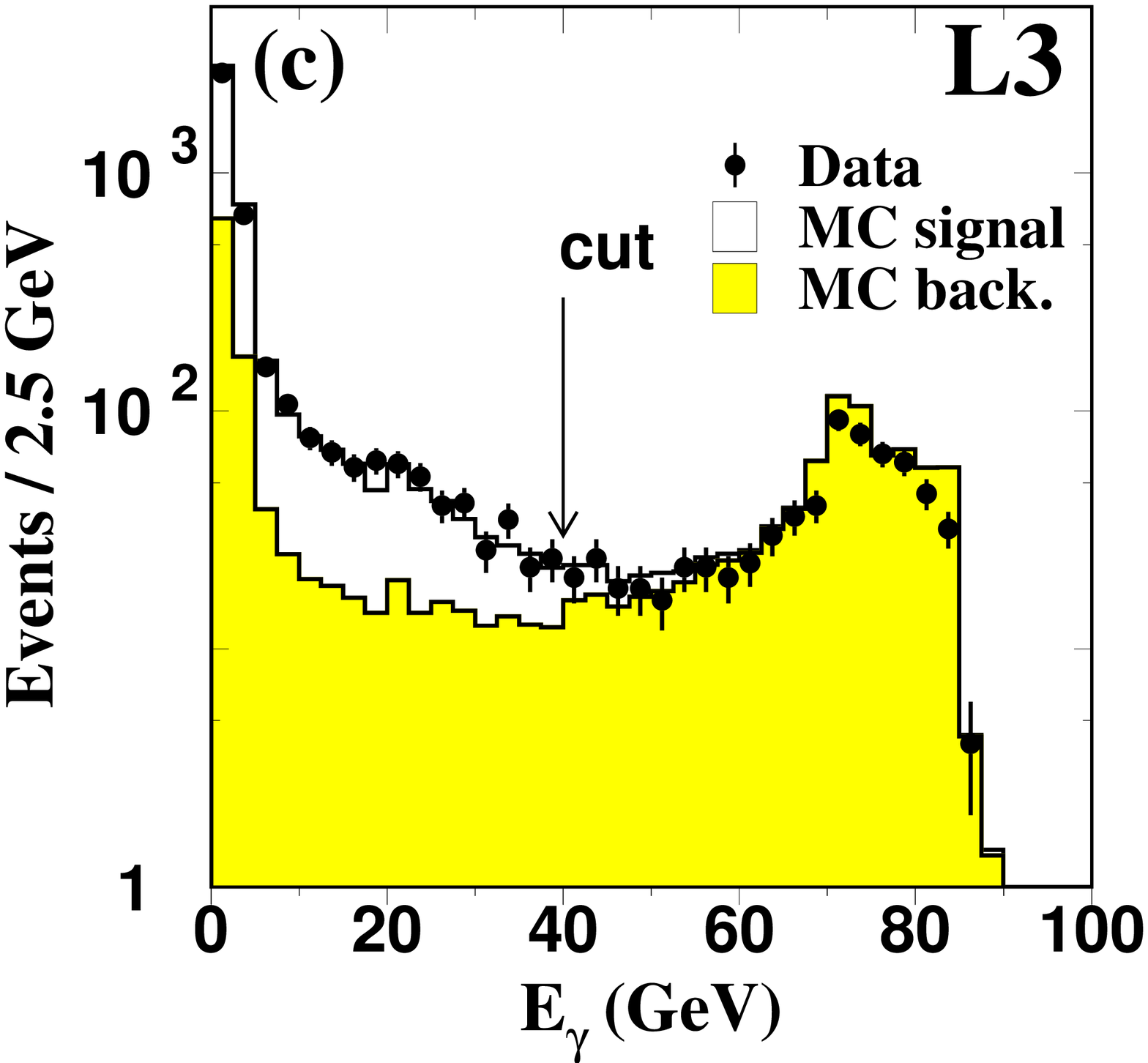}
    \includegraphics[width=8cm]{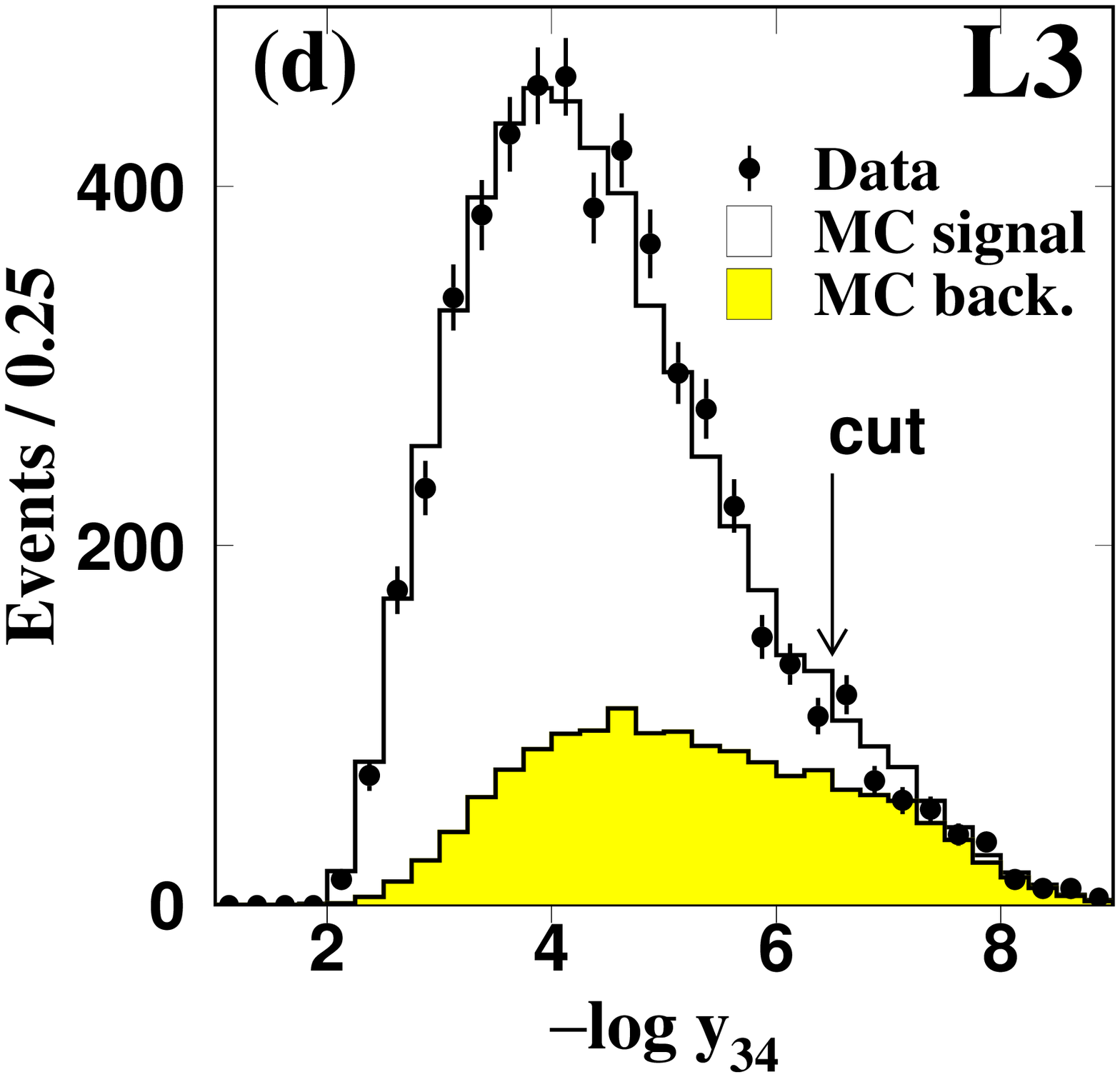}
  \end{center}
  \icaption{Distributions of variables used for the selection of $\epem\ra\ww{}\ra{}\qqqq{}$ events:
(a) normalised visible energy,
(b) event thrust,
(c) energy of the most energetic photon, $E_\gamma$,
(d) jet-resolution parameter, $-\log{y_{34}}$.
In each plot, all other selection criteria are applied.
The arrows indicate the positions of the cuts.
\label{fig:selection}}
\end{figure}

The selection of $\epem\ra\ww\ra\qqqq{}$ events is performed as
follows.  High multiplicity events are selected by requiring more than
20 charged tracks and more than 25 calorimetric clusters.  The visible
energy of the event must satisfy $E_{\text{vis}} > 0.75 \rts{}$.  To
reject the $\epem\ra\qqbar(\gamma)$ background, the event thrust must
be less than 0.88 and the polar angle of the thrust axis,
$\theta_{T}$, has to satisfy $\left|\cos\theta_{T}\right| < 0.95$.
The missing momentum of the event has to be less than 60 \gev{}.
Events containing electrons, muons or photons with energy greater than
20, 20 or 40 \gev{}, respectively, are rejected.  Jets are
reconstructed using the Durham algorithm~\cite{DURHAM}, with a
jet-resolution parameter for which the event changes from a four-jet
into a three-jet topology, $y_{34}$, greater than 0.0015.  Two pairs
of jets are formed, corresponding to two W bosons.  Of the three
combinations, the optimal pairing of jets is chosen as the one with
the smallest mass difference, disregarding the pairing corresponding
to the smallest mass sum.  This algorithm yields the correct
assignment of jets to W bosons for about 70\% of the selected Monte Carlo
events.  Finally, the reconstructed W bosons must have a mass between
40 and 120 \gev{}.  Figure~\ref{fig:selection} shows the distributions
of some selection variables for data and Monte Carlo.

The numbers of events selected by these criteria at different values
of \rts{} are listed in Table~\ref{tab:table1}.  In total, 4919 events
are selected with an average efficiency of 76.3\% and an average
purity of 75.7\%, which do not strongly depend on $\sqrt{s}$.
The background contamination is 18.9\% from the
$\epem\ra{}\qqbar(\gamma)$ process, 4.9\% from the
$\epem\ra{}\textrm{ZZ}$ process and 0.5\% from
W-boson pairs which decay into other final states.

\section*{Analysis of W-Boson Spin Correlations}
The values of \costlep{} and \costhad{} are
determined for each selected event.  The latter is approximated by the absolute value of the
cosine of the angle of the thrust axis of the W-boson decaying into
hadrons, calculated with respect to the W-boson flight direction in
the W-boson rest frame.

\begin{figure}[htbp]\vspace{-5mm}
  \begin{center}
    \includegraphics[height=11cm]{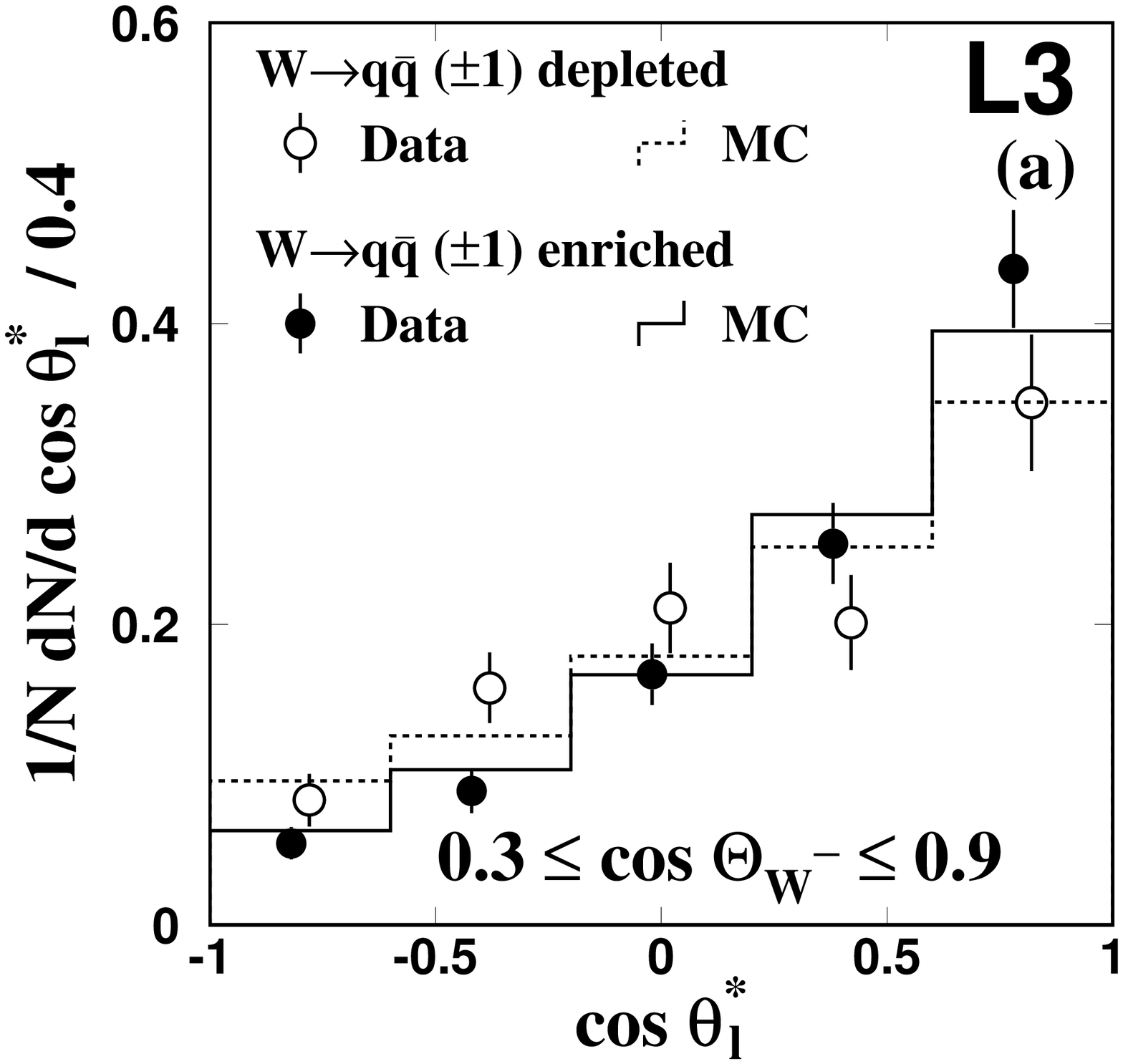}\\ \vspace{-10mm}
    \includegraphics[height=11cm]{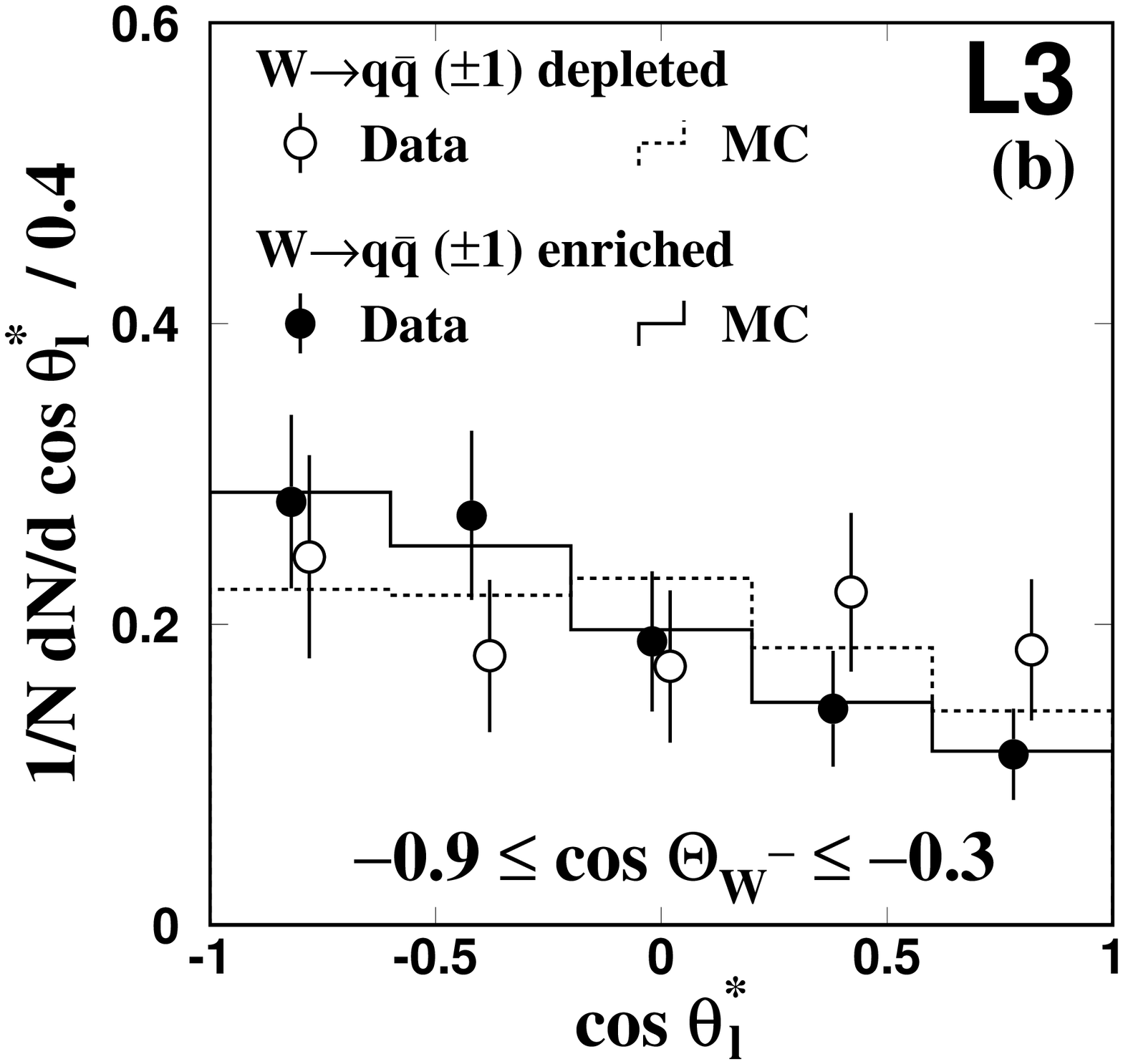}\\ \vspace{-10mm}
  \end{center}
  \caption{Corrected 
$\cos \theta^{*}_{\ell}$ distributions for W $\rightarrow \ell \nu$ decays 
for data and the KORALW Monte Carlo in the intervals 
(a) $0.3 < \cos \Theta_{\text{W}^{-}} < 0.9$, and
(b) $-0.9 < \cos \Theta_{\text{W}^{-}} < -0.3$.
The distributions are shown after classifying the events in two
samples, according to the helicity of the W boson decaying into
hadrons. The first sample is enriched in W bosons in a
transverse-helicity state, the second sample is depleted of W bosons
in a transverse-helicity state.
For clarity, the data points are slightly shifted.
\label{fig:spincorr}}
\end{figure}

The W-pair events are then classified according to the values of
\cosT{} and \costhad{} to build four test samples:
\begin{center}
\begin{tabular}{ll}
$\phantom{-}0.3 < \cosT{} < \phantom{-}0.9$, $\costhad{} < 0.33$ ; &
$\phantom{-}0.3 < \cosT{} < \phantom{-}0.9$, $\costhad{} > 0.66$ ; \\
& \\ $-0.9 < \cosT{} < -0.3$, $\costhad{} < 0.33$ ; & $-0.9 < \cosT{}
< -0.3$, $\costhad{} > 0.66$ . \\
\end{tabular}\\
\end{center}
In each of the samples, the fractions of W-boson helicity states for W
bosons decaying into leptons are obtained from the event
distributions, $\mathrm{d}N/\mathrm{d}\costlep{}$.  For each energy
point the background, as obtained from Monte Carlo simulations, is
subtracted from the data.  The corrected decay angle distributions at different
values of \rts{} are combined into single distributions, shown in
Figure~\ref{fig:spincorr}, which are then fitted with the function in
Equation 1, using \fm{} and \fo{} as the fit parameters.  The fraction
\fp{} is obtained by constraining the sum of all three parameters to
unity.

Finally, the fitted fractions are corrected for the bias which
originates from migration effects due to detector
resolution~\cite{wpol,oldl3}.  Bias correction functions are
determined in each of the investigated \cosT{} and \costhad{} ranges. They vary from 1\% to 15\%.

Several sources of systematic uncertainty are considered, as
summarised in Table~\ref{tab:table3} for the measurement of the
helicity fractions \fm{} and \fo{} for all bins of \cosT{} and
\costhad{}.  All selection cuts are varied over a range of one
standard deviation of the corresponding reconstruction accuracy.  The
corresponding variation of the helicity fractions, corrected for its
statistical component, is taken as systematics.  All fits are repeated
with one bin more or one bin less in the angular distributions.  The
average difference is retained as systematics.  Uncertainties on the
bias and the efficiency corrections are determined by varying the bias
correction function and the efficiencies in each bin by one standard
deviation, as derived from the statistics of the corresponding Monte
Carlo samples.  Additional contamination from the non double-resonant
four-fermion final states is evaluated using the EXCALIBUR Monte
Carlo.  Background levels are varied according to Monte Carlo
statistics for the background processes.

The results of the fits are summarised in Table~\ref{tab:table2}.  The
correlation coefficients of the parameters \fm{} and \fo{} derived
from the fit are shown in Table~\ref{tab:table5}.  W-boson spin
correlations would manifest as sizable differences between the values
of \fm{} and \fo{} measured for the samples depleted of and enriched
in \pmone{} helicity, respectively.  These differences are also listed
in Table~\ref{tab:table2}. There is a general agreement between the
observations and the expectation. The largest difference is observed
for \fm{} in the forward bin, with an observation of $-0.32\pm0.12$ to
be compared with a prediction of $-0.11\pm0.01$. Two consistency tests
are performed. In the first, only data is considered and a confidence
level is calculated for the absence of W-boson spin correlations. This
confidence level is $1.2\%$, which allows to conclude that W-boson
spin correlations are observed. A second test compares the data
with the Standard Model KORALW Monte Carlo. The confidence level for
their compatibility is $34.7\%$.

\section*{Analysis of Decay-Plane Correlations}
Events from the $\epem\ra\ww\ra\lnqq{}$ and $\epem\ra\ww\ra\qqqq{}$
processes are used to study correlations between the W-boson decay
planes.

\begin{figure}[htbp]\vspace{-5mm}
\begin{center}
    \includegraphics[height=12cm]{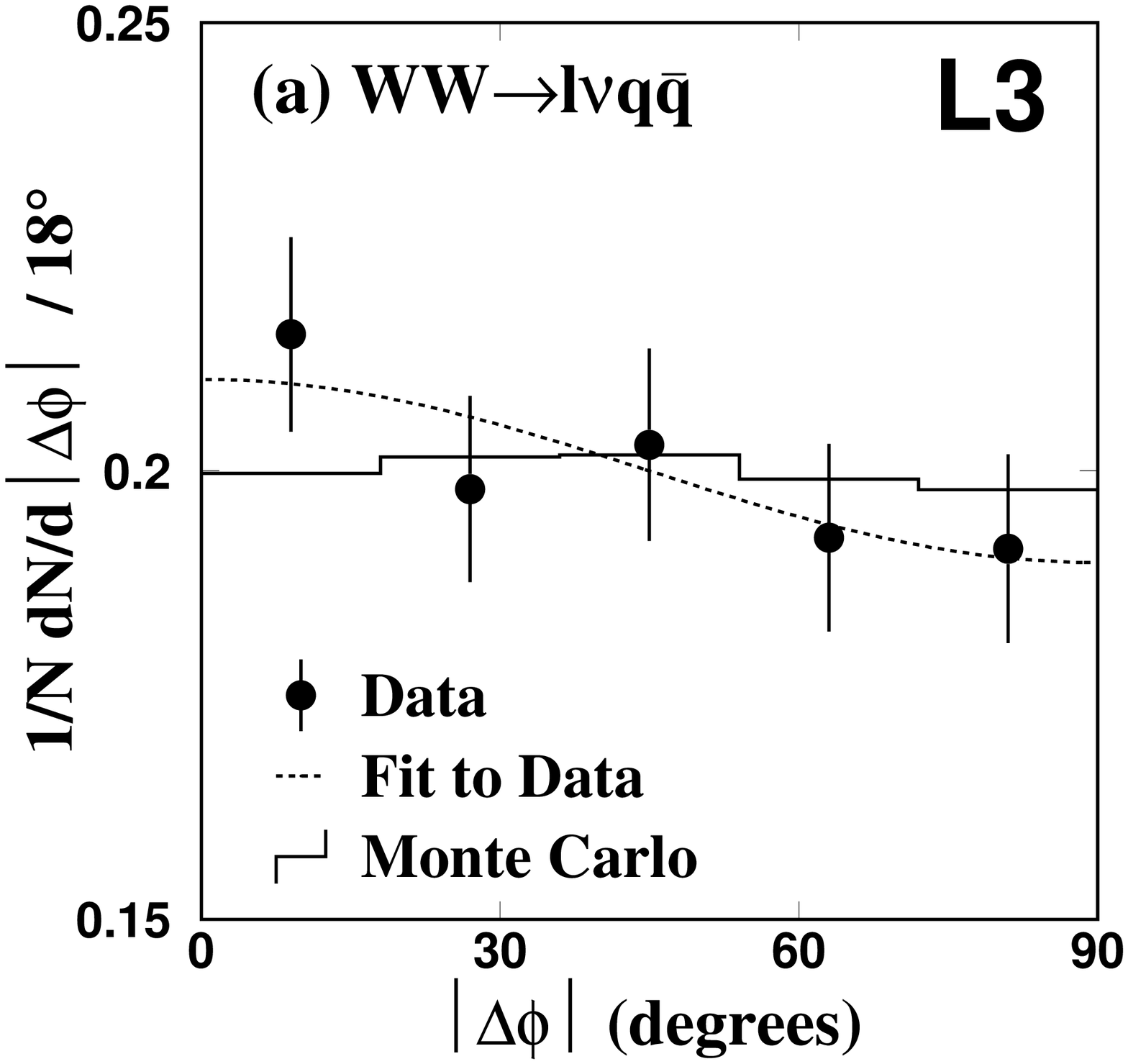}\\ \vspace{-15mm}
    \includegraphics[height=12cm]{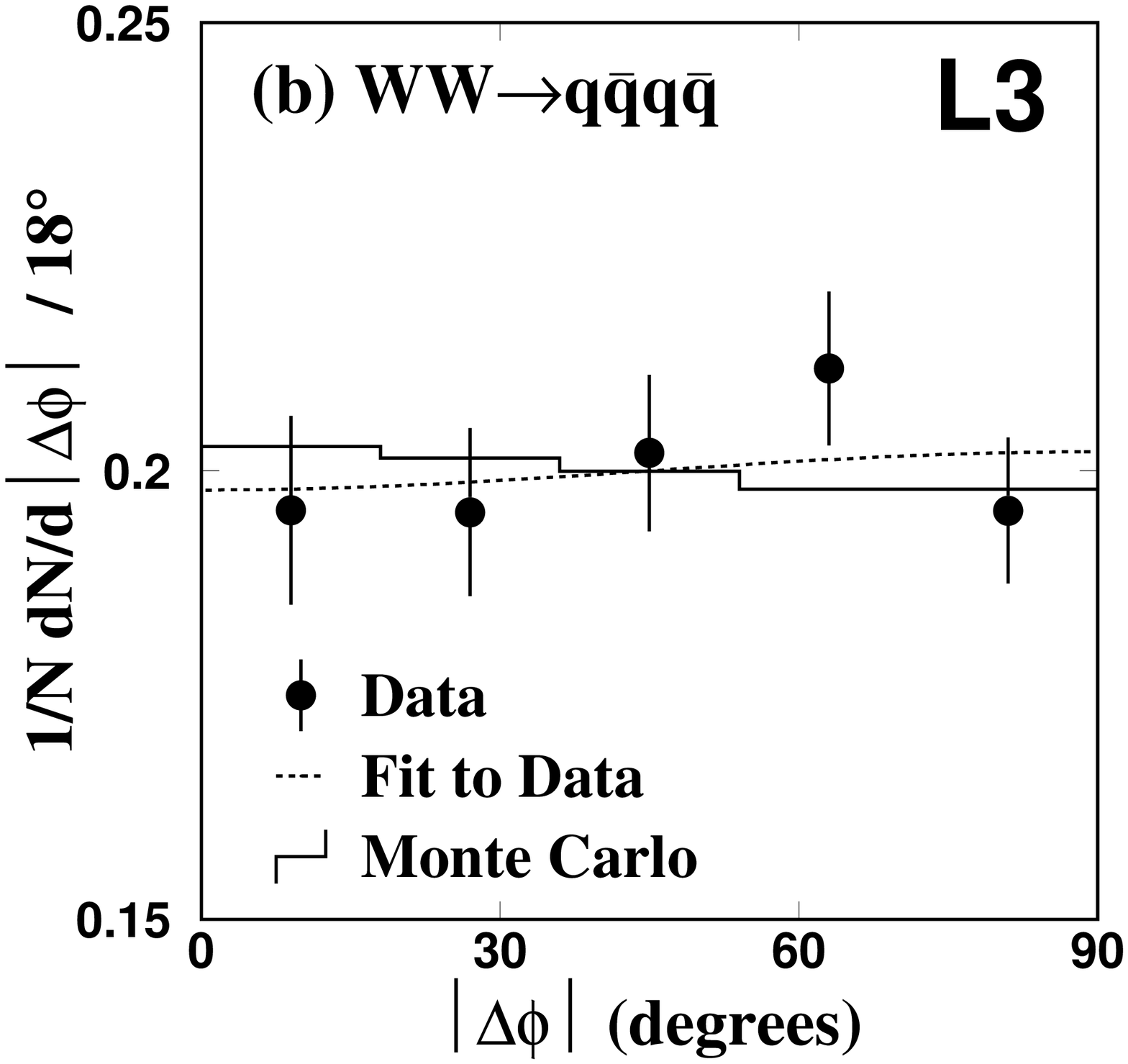}\\ \vspace{-10mm}
\end{center}
\caption{Corrected \dphi{} distributions for (a) $\epem\ra\ww\ra\lnqq{}$ and (b) $\epem\ra\ww\ra\qqqq{}$ events
for data and the KORALW Monte Carlo. The fit results are also shown.
\label{fig:deltaphi}}
\end{figure}

For $\epem\ra\ww\ra\lnqq{}$ events, the neutrino momentum is derived
from the total missing momentum of the event.  The decay plane of the
W boson decaying into leptons is determined from the lepton and the
neutrino directions.  The decay plane of the W boson decaying into
hadrons is determined from its thrust axis in the W-boson rest frame
and the W-boson flight direction.  For $\epem\ra\ww\ra\qqqq{}$ events
the reconstructed jets are boosted into the W-pair rest frame and the
decay planes are determined by the two jets assigned to each
reconstructed W boson.  The angle \dphi{} between the decay planes of
the two W bosons is then calculated.  For each value of \rts{}, the
\dphi{} distribution is corrected for efficiency and background,
taking also into account wrongly-paired four-jet events. The
efficiency corrections are found to be basically independent from the
values of $D$.  The corrected distributions are combined into a single
distribution, shown in Figure~\ref{fig:deltaphi}.  A binned fit for
$D$, using Equation 3, is performed on the normalised distribution.

The same systematic studies are performed as for the W-boson spin
correlations.  Additionally, for $\epem\ra\ww\ra\qqqq{}$ events,
several pairing algorithms are used as a cross-check.  The largest
difference in the fit result between the pairing methods is taken as
systematic uncertainty.  To reproduce the measured four jet event rate
of the $\epem\ra\qqbar(\gamma)$ background~\cite{paper218}, the
corresponding Monte Carlo is scaled by $+10\%$.  Half of the effect is
retained as systematics.  Table~\ref{tab:table4} summarises the
systematic uncertainties in the measurement of the decay-plane
correlation parameter $D$.

The resulting value of $D$ for $\epem\ra\ww\ra\lnqq{}$ events is found
to be $0.051 \pm 0.033 \pm 0.019$, where the first uncertainty is
statistical and the second systematic.  It is in agreement with the
Standard Model prediction from the KORALW Monte Carlo of $D = 0.006
\pm 0.004$, where the error reflects the Monte Carlo statistics.  For
$\epem\ra\ww\ra\qqqq{}$ events, $D$ is found to be $-0.016 \pm 0.028
\pm 0.016$, in agreement with the KORALW prediction of $D = 0.013 \pm
0.003$.  Combining the two decay channels, a value $D = 0.012 \pm
0.021 \pm 0.012$ is found in data, in agreement with the combined
value from the Standard Model KORALW Monte Carlo of $D = 0.010 \pm
0.002$.

\section*{Summary}
In conclusion, this study completes our investigation of W-boson
polarisation. A new technique, which is promising for future
high-luminosity electron-positron colliders, is deployed and allows to
establish the existence of W-boson spin correlations. Their magnitude
is found to agree with the Standard Model predictions. In addition,
W-boson decay-plane correlations are studied for the first time, and
no large deviations with respect to the Standard Model predictions,
which could hint to New Physics, are observed.

%
%
\bibliographystyle{/l3/paper/biblio/l3stylem}

%
%
\newpage
\typeout{   }     
\typeout{Using author list for paper 281 -  }
\typeout{$Modified: Jul 15 2001 by smele $}
\typeout{!!!!  This should only be used with document option a4p!!!!}
\typeout{   }
%
%
%
%
%
%

\newcount\tutecount  \tutecount=0
\def\tutenum#1{\global\advance\tutecount by 1 \xdef#1{\the\tutecount}}
\def\tute#1{$^{#1}$}
\tutenum\aachen            
\tutenum\nikhef            
\tutenum\mich              
\tutenum\lapp              
\tutenum\basel             
\tutenum\lsu               
\tutenum\beijing           
\tutenum\bologna           
\tutenum\tata              
\tutenum\ne                
\tutenum\bucharest         
\tutenum\budapest          
\tutenum\mit               
\tutenum\panjab            
\tutenum\debrecen          
\tutenum\dublin            
\tutenum\florence          
\tutenum\cern              
\tutenum\wl                
\tutenum\geneva            
\tutenum\hefei             
\tutenum\lausanne          
\tutenum\lyon              
\tutenum\madrid            
\tutenum\florida           
\tutenum\milan             
\tutenum\moscow            
\tutenum\naples            
\tutenum\cyprus            
\tutenum\nymegen           
\tutenum\caltech           
\tutenum\perugia           
\tutenum\peters            
\tutenum\cmu               
\tutenum\potenza           
\tutenum\prince            
\tutenum\riverside         
\tutenum\rome              
\tutenum\salerno           
\tutenum\ucsd              
\tutenum\sofia             
\tutenum\korea             
\tutenum\purdue            
\tutenum\psinst            
\tutenum\zeuthen           
\tutenum\eth               
\tutenum\hamburg           
\tutenum\taiwan            
\tutenum\tsinghua          

{
\parskip=0pt
\noindent
{\bf The L3 Collaboration:}
\ifx\selectfont\undefined
 \baselineskip=10.8pt
 \baselineskip\baselinestretch\baselineskip
 \normalbaselineskip\baselineskip
 \ixpt
\else
 \fontsize{9}{10.8pt}\selectfont
\fi
\medskip
\tolerance=10000
\hbadness=5000
\raggedright
\hsize=162truemm\hoffset=0mm
\def\r{\rlap,}
\noindent

P.Achard\r\tute\geneva\ 
O.Adriani\r\tute{\florence}\ 
M.Aguilar-Benitez\r\tute\madrid\ 
J.Alcaraz\r\tute{\madrid}\ 
G.Alemanni\r\tute\lausanne\
J.Allaby\r\tute\cern\
A.Aloisio\r\tute\naples\ 
M.G.Alviggi\r\tute\naples\
H.Anderhub\r\tute\eth\ 
V.P.Andreev\r\tute{\lsu,\peters}\
% ...
\newpage
\begin{list}{A}{\itemsep=0pt plus 0pt minus 0pt\parsep=0pt plus 0pt minus 0pt
                \topsep=0pt plus 0pt minus 0pt}
\item[\aachen]
 III. Physikalisches Institut, RWTH, D-52056 Aachen, Germany$^{\S}$
\item[\nikhef] National Institute for High Energy Physics, NIKHEF, 
     and University of Amsterdam, NL-1009 DB Amsterdam, The Netherlands
\item[\mich] University of Michigan, Ann Arbor, MI 48109, USA
\item[\lapp] Laboratoire d'Annecy-le-Vieux de Physique des Particules, 
     LAPP,IN2P3-CNRS, BP 110, F-74941 Annecy-le-Vieux CEDEX, France
\item[\basel] Institute of Physics, University of Basel, CH-4056 Basel,
     Switzerland
\item[\lsu] Louisiana State University, Baton Rouge, LA 70803, USA
\item[\beijing] Institute of High Energy Physics, IHEP, 
  100039 Beijing, China$^{\triangle}$ 
\item[\bologna] University of Bologna and INFN-Sezione di Bologna, 
     I-40126 Bologna, Italy
\item[\tata] Tata Institute of Fundamental Research, Mumbai (Bombay) 400 005, India
\item[\ne] Northeastern University, Boston, MA 02115, USA
\item[\bucharest] Institute of Atomic Physics and University of Bucharest,
     R-76900 Bucharest, Romania
\item[\budapest] Central Research Institute for Physics of the 
     Hungarian Academy of Sciences, H-1525 Budapest 114, Hungary$^{\ddag}$
\item[\mit] Massachusetts Institute of Technology, Cambridge, MA 02139, USA
\item[\panjab] Panjab University, Chandigarh 160 014, India.
\item[\debrecen] KLTE-ATOMKI, H-4010 Debrecen, Hungary$^\P$
\item[\dublin] Department of Experimental Physics,
  University College Dublin, Belfield, Dublin 4, Ireland
\item[\florence] INFN Sezione di Firenze and University of Florence, 
     I-50125 Florence, Italy
\item[\cern] European Laboratory for Particle Physics, CERN, 
     CH-1211 Geneva 23, Switzerland
\item[\wl] World Laboratory, FBLJA  Project, CH-1211 Geneva 23, Switzerland
\item[\geneva] University of Geneva, CH-1211 Geneva 4, Switzerland
\item[\hefei] Chinese University of Science and Technology, USTC,
      Hefei, Anhui 230 029, China$^{\triangle}$
\item[\lausanne] University of Lausanne, CH-1015 Lausanne, Switzerland
\item[\lyon] Institut de Physique Nucl\'eaire de Lyon, 
     IN2P3-CNRS,Universit\'e Claude Bernard, 
     F-69622 Villeurbanne, France
\item[\madrid] Centro de Investigaciones Energ{\'e}ticas, 
     Medioambientales y Tecnol\'ogicas, CIEMAT, E-28040 Madrid,
     Spain${\flat}$ 
\item[\florida] Florida Institute of Technology, Melbourne, FL 32901, USA
\item[\milan] INFN-Sezione di Milano, I-20133 Milan, Italy
\item[\moscow] Institute of Theoretical and Experimental Physics, ITEP, 
     Moscow, Russia
\item[\naples] INFN-Sezione di Napoli and University of Naples, 
     I-80125 Naples, Italy
\item[\cyprus] Department of Physics, University of Cyprus,
     Nicosia, Cyprus
\item[\nymegen] University of Nijmegen and NIKHEF, 
     NL-6525 ED Nijmegen, The Netherlands
\item[\caltech] California Institute of Technology, Pasadena, CA 91125, USA
\item[\perugia] INFN-Sezione di Perugia and Universit\`a Degli 
     Studi di Perugia, I-06100 Perugia, Italy   
\item[\peters] Nuclear Physics Institute, St. Petersburg, Russia
\item[\cmu] Carnegie Mellon University, Pittsburgh, PA 15213, USA
\item[\potenza] INFN-Sezione di Napoli and University of Potenza, 
     I-85100 Potenza, Italy
\item[\prince] Princeton University, Princeton, NJ 08544, USA
\item[\riverside] University of Californa, Riverside, CA 92521, USA
\item[\rome] INFN-Sezione di Roma and University of Rome, ``La Sapienza",
     I-00185 Rome, Italy
\item[\salerno] University and INFN, Salerno, I-84100 Salerno, Italy
\item[\ucsd] University of California, San Diego, CA 92093, USA
\item[\sofia] Bulgarian Academy of Sciences, Central Lab.~of 
     Mechatronics and Instrumentation, BU-1113 Sofia, Bulgaria
\item[\korea]  The Center for High Energy Physics, 
     Kyungpook National University, 702-701 Taegu, Republic of Korea
\item[\purdue] Purdue University, West Lafayette, IN 47907, USA
\item[\psinst] Paul Scherrer Institut, PSI, CH-5232 Villigen, Switzerland
\item[\zeuthen] DESY, D-15738 Zeuthen, Germany
\item[\eth] Eidgen\"ossische Technische Hochschule, ETH Z\"urich,
     CH-8093 Z\"urich, Switzerland
\item[\hamburg] University of Hamburg, D-22761 Hamburg, Germany
\item[\taiwan] National Central University, Chung-Li, Taiwan, China
\item[\tsinghua] Department of Physics, National Tsing Hua University,
      Taiwan, China
\item[\S]  Supported by the German Bundesministerium 
        f\"ur Bildung, Wissenschaft, Forschung und Technologie
\item[\ddag] Supported by the Hungarian OTKA fund under contract
numbers T019181, F023259 and T037350.
\item[\P] Also supported by the Hungarian OTKA fund under contract
  number T026178.
\item[$\flat$] Supported also by the Comisi\'on Interministerial de Ciencia y 
        Tecnolog{\'\i}a.
\item[$\sharp$] Also supported by CONICET and Universidad Nacional de La Plata,
        CC 67, 1900 La Plata, Argentina.
\item[$\triangle$] Supported by the National Natural Science
  Foundation of China.
\end{list}
}
\vfill


\newpage

\begin{table}[htbp]
 \begin{center}
  \begin{tabular}{|c|c|c|c|c|c|c|c|}
    \hline
    $\langle\rts{}\rangle$ [\gev{}]& 188.6 & 191.6 & 195.5 & 199.5 & 201.8 & 205.9 \\
    \hline
    Integrated luminosity [\pb{}] & 176.8 & \pho{}29.8 & \pho{}84.1 & \pho{}83.3 & \pho{}37.2 & 218.1 \\
    \hline
    $\epem\ra\enqq{}$ & \pho{}293 & \pho{}59 &  133 &  110 & \pho{}56 & \pho{}355 \\
    $\epem\ra\mnqq{}$ & \pho{}255 & \pho{}43 &  110 & \pho{}99 & \pho{}59 & \pho{}289  \\
    $\epem\ra\qqqq{}$ & 1447 & 224 & 640 & 683 & 269 & 1656 \\
    \hline
    \end{tabular}
    \icaption{Average centre-of-mass energies with corresponding integrated luminosities and numbers of selected events.
    \label{tab:table1}}
  \end{center}
\end{table}

\begin{table}[htbp]
  \begin{center}
    \begin{tabular}{|l|c|c|c|c|c|c|c|c|}\hline
          & \multicolumn{4}{|c|}{$0.3 < \cosT{} < 0.9$} & \multicolumn{4}{|c|}{$-0.9 < \cosT{} < -0.3$} \\
	  \cline{2-9}
	  & \multicolumn{2}{|c|}{\pmone{} depleted} & \multicolumn{2}{|c|}{\pmone{} enriched} & \multicolumn{2}{|c|}{\pmone{} depleted} & \multicolumn{2}{|c|}{\pmone{} enriched} \\
	  \cline{2-9}
	  & \fm{} & \fo{} & \fm{} & \fo{} & \fm{} & \fo{} & \fm{} & \fo{} \\
        \hline
        Selection 	                & 0.034 & 0.045 & 0.010 &           $<$ 0.001 & 0.053 & 0.048 & 0.050 & 0.087 \\ 
        Fit binning                     & 0.027 & 0.021 & 0.030 & $\phantom{<}$ 0.057 & 0.078 & 0.097 & 0.020 & 0.051 \\
        Bias correction                 & 0.018 & 0.026 & 0.010 & $\phantom{<}$ 0.013 & 0.031 & 0.055 & 0.016 & 0.034 \\
	Efficiency correction		& 0.003 & 0.001 & 0.005 & $\phantom{<}$ 0.004 & 0.004 & 0.006 & 0.002 & 0.005 \\
        Four fermions                   & 0.001 & 0.006 & 0.006 & $\phantom{<}$ 0.003 & 0.012 & 0.024 & 0.003 & 0.026 \\
        Background levels               & 0.006 & 0.008 & 0.004 & $\phantom{<}$ 0.008 & 0.022 & 0.024 & 0.014 & 0.019 \\
        \hline
        Total                           & 0.048 & 0.057 & 0.034 & $\phantom{<}$ 0.059 & 0.102 & 0.126 & 0.058 & 0.111 \\
        \hline
    \end{tabular}
    \icaption{Systematic uncertainties on the measurement of \fm{} and
    \fo{} for leptonic W-boson decays in bins of \cosT{} and for
    samples depleted of and enriched in the \pmone{} helicity. The
    statistical component of these uncertainties is removed.
    \label{tab:table3}}
  \end{center}
\end{table}

\begin{table}[htbp]
  \begin{center}
    \begin{tabular}{|c|ccc|}\hline
	& \multicolumn{3}{|c|}{$0.3 < \cosT{} < 0.9$: $(\lambda_{\Wm{}},\lambda_{\Wp{}}) = (-1,+1)$ enriched}\\
        & \fm{} & \fp{} & \fo{} \\
	\hline
	Data \pmone{} depleted &  $\phantom{-}0.521\pm0.086\pm0.048$ & $0.121\pm0.058\pm0.035$ & $0.358\pm0.116\pm0.057$ \\
	Data \pmone{} enriched &  $\phantom{-}0.839\pm0.057\pm0.034$ & $0.087\pm0.034\pm0.036$ & $0.074\pm0.072\pm0.059$ \\
	Difference             &            $-0.318\pm0.103\pm0.059$ & $0.034\pm0.067\pm0.050$ & $0.284\pm0.137\pm0.082$ \\
	\hline
	MC \pmone{} depleted   &  $\phantom{-}0.670\pm0.008$ & $0.148\pm0.006$ & $0.182\pm0.012$ \\
	MC \pmone{} enriched   &  $\phantom{-}0.781\pm0.007$ & $0.091\pm0.004$ & $0.128\pm0.009$ \\
	Difference             &            $-0.111\pm0.011$ & $0.057\pm0.007$ & $0.054\pm0.015$ \\
	\hline
	\hline
	& \multicolumn{3}{|c|}{$-0.9 < \cosT{} < -0.3$: $(\lambda_{\Wm{}},\lambda_{\Wp{}}) = (0,0)$ enriched }\\
        & \fm{} & \fp{} & \fo{} \\
	\hline
	Data \pmone{} depleted & $0.387\pm0.136\pm0.102$ &  $\phantom{-}0.398\pm0.121\pm0.061$ &  $\phantom{-}0.215\pm0.316\pm0.126$ \\
	Data \pmone{} enriched & $0.152\pm0.082\pm0.058$ &  $\phantom{-}0.530\pm0.101\pm0.087$ &  $\phantom{-}0.318\pm0.215\pm0.111$ \\
	Difference             & $0.235\pm0.159\pm0.117$ &            $-0.132\pm0.158\pm0.107$ &            $-0.103\pm0.382\pm0.168$ \\
	\hline
	MC \pmone{} depleted   & $0.186\pm0.014$ &  $\phantom{-}0.351\pm0.016$ & $0.463\pm0.023$ \\
	MC \pmone{} enriched   & $0.158\pm0.014$ &  $\phantom{-}0.532\pm0.017$ & $0.310\pm0.026$ \\
	Difference             & $0.028\pm0.020$ &            $-0.181\pm0.023$ & $0.153\pm0.035$ \\
      \hline
    \end{tabular}
\icaption{The W-boson helicity fractions measured for different intervals of \cosT{}.
The results are shown for different subsamples depleted of or enriched in helicity \pmone{}.
The corresponding helicity fractions in the Standard Model,
as implemented in the KORALW Monte Carlo program,
are also given with their statistical uncertainties.
    \label{tab:table2} }
  \end{center}
\end{table}

\begin{table}[htbp]
  \begin{center}
    \begin{tabular}{|c|c|c|c|c|}\hline
 & \multicolumn{4}{|c|}{\fm{} $-$ \fo{} correlation coefficient}\\
\cline{2-5}
 & \multicolumn{2}{|c|}{Data} & \multicolumn{2}{|c|}{Monte Carlo} \\
\cline{2-5}
 & \pmone{} depleted & \pmone{} enriched & \pmone{} depleted & \pmone{} enriched \\
\hline
$0.3 < \cosT{} < 0.9$ 	& $-92\%$ & $-93\%$ & $-88\%$ & $-91\%$ \\
\hline
$-0.9 < \cosT{} < -0.3$	& $-76\%$ & $-77\%$ & $-80\%$ & $-78\%$ \\
\hline
    \end{tabular}
\icaption{Correlation coefficients between the fit parameters \fm{} and \fo{}.
    \label{tab:table5} }
  \end{center}
\end{table}

\begin{table}[htbp]
  \begin{center}
    \begin{tabular}{|l|c|c|}\hline
                                        & $\epem\ra\ww\ra\lnqq{}$ & $\epem\ra\ww\ra\qqqq$ \\
        \hline
        Selection                       & 0.013 & 0.007 \\
        Fit binning                     & 0.012 & 0.007 \\
	Efficiency correction		& 0.001 & 0.002 \\
        Four fermions                   & 0.005 & --- \\
        Background levels               & 0.002 & 0.007 \\
	Jet pairing 			& --- & 0.011 \\
        \hline
        Total                           & 0.019 & 0.016  \\
        \hline
    \end{tabular}
    \icaption{Systematic uncertainties on the measurement of the
    decay-plane correlation parameter $D$. The statistical component
    of these uncertainties is removed.
    \label{tab:table4}}
  \end{center}
\end{table}


\end{document}